\documentclass[aps,prd,a4paper,twocolumn,amsmath,showpacs,superscriptaddress,nofootinbib,preprintnumbers]{revtex4-1}

\usepackage{verbatim}
\usepackage[T1]{fontenc}
\usepackage[utf8]{inputenc}
\usepackage[american]{babel}
\usepackage{epsfig}
\usepackage{graphicx}
\usepackage{booktabs}
\usepackage{multirow}
\usepackage{dcolumn}
\usepackage{amsmath}
\usepackage{mathtools}
\usepackage{amsfonts}
\usepackage{amssymb}
\usepackage{epstopdf}
\usepackage{bm}
\usepackage{siunitx}
\usepackage{braket}
\usepackage{enumitem}
\usepackage{soul}
\usepackage[table]{xcolor}
\usepackage{color}
\usepackage{transparent}
\usepackage{pifont}

\definecolor{navyblue}{rgb}{0.0, 0.0, 0.5}
\definecolor{royalblue}{rgb}{0.25, 0.41, 0.88}
\definecolor{cadmiumgreen}{rgb}{0.0, 0.42, 0.24}
\definecolor{blue-violet}{rgb}{0.54, 0.17, 0.89}
\definecolor{darkviolet}{rgb}{0.58, 0.0, 0.83}
\definecolor{orange(colorwheel)}{rgb}{1.0, 0.5, 0.0}

\usepackage{hyperref}
\hypersetup{
    colorlinks=true, % false: boxed links; true: colored links
    linkcolor=royalblue, % color of internal links (change box color with linkbordercolor)
    citecolor=magenta}

\newcommand\ee{\end{equation}}
\newcommand\be{\begin{equation}}
\newcommand\eea{\end{eqnarray}}
\newcommand\bea{\begin{eqnarray}}

\usepackage{booktabs}
\usepackage{multirow}
\usepackage{dcolumn}
\usepackage{colortbl}
\newcommand\vertsp{\rule[-2mm]{1mm}{0mm} &}
\newcommand\horsp{\rule[-1.5mm]{0mm}{4.125mm}}
\newcommand\morehorsp{\rule[-2.25mm]{0mm}{6mm}}

\definecolor{magenta(process)}{rgb}{1.0, 0.0, 0.56}

\definecolor{darkspringgreen}{rgb}{0.09, 0.45, 0.27}

\definecolor{royalblue(web)}{rgb}{0.25, 0.41, 0.88}
\begin{document}

\title{Cornering the Planck $A_{lens}$ tension with future CMB data.}  

\author{Fabrizio Renzi}
\email{fabrizio.renzi@roma1.infn.it}
\affiliation{Physics Department and INFN, Universit\`a di Roma ``La Sapienza'', Ple Aldo Moro 2, 00185, Rome, Italy} 

\author{Eleonora Di Valentino}
\email{eleonora.divalentino@manchester.ac.uk}
\affiliation{Jodrell Bank Center for Astrophysics, School of Physics and Astronomy, University of Manchester, Oxford Road, Manchester, M13 9PL, UK}

\author{Alessandro Melchiorri}
\email{alessandro.melchiorri@roma1.infn.it}
\affiliation{Physics Department and INFN, Universit\`a di Roma ``La Sapienza'', Ple Aldo Moro 2, 00185, Rome, Italy} 

\date{\today}

\preprint{}
\begin{abstract}
The precise measurements of Cosmic Microwave Background Anisotropy angular power spectra made by the Planck satellite show an anomalous value for the lensing amplitude, defined by the parameter $A_{lens}$, at about $2$ standard deviations ($2.6$ standard deviations when cosmic shear data is included). Moreover, considering $A_{lens}$ brings the values of the cosmological parameters determined by Planck in better agreement with those found by pre-Planck datasets. In this paper, after discussing the current status of the anomaly, we quantify the potential of future CMB measurements in confirming/falsifying the $A_{lens}$ tension. We find that a space-based experiment as LiteBIRD could falsify the current $A_{lens}$ tension at the level of $5$ standard deviations. Similar constraints can be achieved by a Stage-III experiment assuming an external prior on the reionization optical depth of $\tau=0.055\pm0.010$ as already provided by the Planck satellite. A Stage-IV experiment could further test the $A_{lens}$ tension at the level of $10$ standard deviations.  A comparison between temperature and polarization measurements made at different frequencies could further identify possible systematics responsible for $A_{lens}>1$. We show that, in the case of the CMB-S4 experiment, polarization data alone  will have the potential of falsifying the current 
$A_{lens}$ anomaly at more than five standard deviation and to strongly bound its frequency dependence. We also evaluate the future constraints on a possible scale dependence for $A_{lens}$. 
\end{abstract}

\maketitle

%%%%%%%%%%%%%%%%%%%%%
\section{Introduction}

The precise measurements of Cosmic Microwave Background (CMB) anisotropies made by the Planck satellite \cite{planck2015} have provided a wonderful confirmation of the standard cosmological model of structure formation based on inflation, dark matter and a cosmological constant. The predictions of acoustic oscillations in the CMB anisotropy angular power spectra have been fully confirmed with unprecedented accuracy. 

Nonetheless few, interesting, tensions are emerging hinting to systematics and/or possible extensions to the standard scenario (see e.g. \cite{plancklike,planckshift,bennett,hulensing}).

The most relevant anomaly, at least from the statistical point of view, concerns the amount of lensing in the CMB angular power spectra.  Gravitational lensing slightly redistributes the photon paths from the last scattering surface, smoothing the acoustic oscillations in the CMB anisotropy and polarization power spectra (see \cite{seljak}).

The amount of smearing due to CMB lensing, once the cosmological parameters are fixed, can be computed with great accuracy (see e.g. \cite{chalew}) and the effect is included in all current parameter analyses. In \cite{calens} a phenomenological parameter, $A_{lens}$, was introduced that essentially rescales the lensing amplitude in the CMB spectra. This parameter has, in principle, no physical meaning and is mainly used as an effective parameter for testing theoretical assumptions and systematics. However, the value of this parameter from the latest Planck analysis of \cite{plancknewtau} is $A_{lens}=1.15_{-0.12}^{+0.13}$ at $95 \%$ c.l., i.e. about $2.3$ $\sigma$ larger than the expected value with a significant impact on parameter extraction.

Indeed, the inclusion of $A_{lens}$ in the analysis shifts the constraints derived from Planck data on several cosmological parameters. Interestingly, some tension exists between the cosmological parameters derived from a combination of pre-Planck datasets and those obtained by the Planck satellite (see Table I in \cite{calabresenew} and discussion in \cite{planckshift,bennett}). As noted in \cite{planckshift,bennett} and as we report in Appendix I of this paper, the inclusion of $A_{lens}$ significantly reduces this tension.

Moreover, lensing in the CMB spectra is crucial in constraining neutrino masses. A larger value for $A_{lens}$, if not accounted for, could produce biased bounds on neutrino masses, stronger than those that realistically could be reached with the Planck specifications and experimental noise. Indeed, from simulated Planck angular spectra (assuming a neutrino mass of $\Sigma m_{\nu} \le 0.06$ eV), one would expect a limit on the sum of neutrino masses of $\Sigma m_{\nu} \le 0.59$ eV at $95 \%$ c.l., while the current limit from real Planck data is much stronger, at the level of $\Sigma m_{\nu} \le 0.34$ eV at $95 \%$ c.l. (see \cite{plancknewtau}). These stronger than expected neutrino mass bounds from Planck are connected to the $2.3$ $\sigma$ $A_{lens}$ tension and should be treated with great care.

Finally, $A_{lens}$ anti-correlates with the amplitude of r.m.s. matter density fluctuations on $8 h^{-1} Mpc$ scales, the so-called $\sigma_8$ parameter. Allowing $A_{lens}$ to vary brings indeed the constraints on the $S_8=\sigma_8 (\Omega_m/0.3)^{0.5}$ parameter from $S_8=0.852\pm0.018$ at $68 \%$ C.L. to $S_8=0.808\pm0.034$, in better agreement with the constraints derived from cosmic shear data from the KiDS-450 \cite{kids} and DES \cite{deswl1,deswl2} surveys.

While $A_{lens}$ seems to solve several current tensions, there are at least two puzzling aspects of the $A_{lens}$ anomaly that should suggest some caution. First of all, there is no easy theoretical way to accommodate a value of $A_{lens}$ {\it larger} than expected, even in an extended parameter space (see e.g. \cite{papero0,papero,paper1}). Proposals that can give a theoretical explanation to the $A_{lens}$ anomaly include, for example, modified gravity \cite{modgrav}, running of the running of the spectral index \cite{runrun}, closed universes \cite{ratra}, and compensated baryon isocurvature perturbations \cite{CIP,CIP2}. These explanations are certainly all rather exotic and would hint for a significant change in the standard scenario. The second point is that an anomalous $A_{lens}$ value, if related to lensing, must show up also in the CMB lensing measurements based on the trispectrum analysis of the Planck temperature and polarization maps. However Planck CMB lensing is in perfect agreement with the standard expectations. Combining the Planck angular power spectra with the CMB lensing yields $A_{lens}=1.025^{+0.051}_{-0.058}$ \cite{planck2015}, in agreement with the standard value even if at the price of an higher $\chi^2$ value due to the relative inconsistency between the two datasets. This fact in practice, even if based on the assumption of $\Lambda$CDM, disfavors the hypothesis of $A_{lens}>1$ due to gravitational lensing.

These two aspects could suggest that the $A_{lens}$ anomaly is related to some systematics in the data. However, the anomaly survived the scrutiny of two Planck data releases and hints for its presence have already been reported, albeit at small statistical level, in pre-Planck data (see e.g. \cite{tail}). 

It is therefore timely to investigate the potential of future CMB experiments to confirm and/or rule out the $A_{lens}$ anomaly. Several ground and space-based  experiments are indeed proposed or expected in the next years that will sample the small scale region of the CMB angular spectrum. At the same time it is important to scrutinize the ability of these experiments in detecting a possible scale dependence of the effect. 

This is indeed the goal of the present paper. While this kind of analysis is straightforward, none of the several recent papers that forecasted the ability of future experiments in constraining cosmological parameters (see e.g. \cite{core1,erminia,stage4}), as far as we know, considered the $A_{lens}$ parameter.

In the next Section we briefly discuss the current status of the $A_{lens}$ tension. In Section III we describe the data analysis method adopted for our forecasts. In Section IV we show the obtained results and in Section V we present our conclusions.

\section{Current status of the $A_{lens}$ anomaly.}

\begin{table*}[!hbtp]
\begin{center}
\begin{tabular}{lcccc}
\toprule
\horsp
Parameter & WMAP9 & Planck TTTEEE & Planck TTTEEE& Planck TTTEEE\\
 & +ACT+SPT &  & ($A_{lens}$)& +WL ($A_{lens}$)\\
\hline
\morehorsp
$100\Omega_bh^2$&$2.242 \pm 0.032$ &$2.222 \pm 0.015$ $[0.56]$&$2.239\pm0.017$ $[0.08]$&$2.245\pm0.017$ $[0.08]$\\
\morehorsp
$100\Omega_ch^2$&$11.34 \pm 0.36$&$12.03 \pm 0.14$ $[1.79]$&$11.87\pm0.16$ $[1.34]$&$11.78\pm0.15$ $[1.13]$\\
\morehorsp
$10^{4}\theta_{MC}$&$104.24 \pm 0.10$&$104.069 \pm 0.032$ $[1.63]$&$104.09\pm0.033$ $[1.42]$&$104.10\pm0.033$ $[1.32]$\\
\morehorsp
$n_s$&$0.9638 \pm 0.0087$&$0.9626 \pm 0.0044$ $[0.12]$&$0.9675\pm0.0049$ $[0.37]$&$0.9697\pm0.0047$ $[0.59]$\\
\morehorsp
$\Omega_{\Lambda}$&$0.723 \pm 0.019$&$0.6812 \pm 0.0086$ $[2.00]$&$0.6920\pm0.0096$ $[1.46]$&$0.6974\pm0.0089$ $[1.22]$\\
\morehorsp
$\Omega_{m}$&$0.277 \pm 0.019$&$0.3188 \pm 0.0086$ $[2.00]$&$0.3080\pm0.0096$ $[1.46]$&$0.3026\pm0.0089$ $[1.22]$\\
\morehorsp
$\sigma_8$&$0.780 \pm 0.017$&$0.8212 \pm 0.0086$ $[2.16]$&$0.806\pm0.017$ $[1.08]$&$0.797\pm0.016$ $[0.73]$\\
\morehorsp
$t_0$ [Gyrs] &$13.787\pm0.057$&$13.822 \pm 0.025$ $[0.56]$&$13.790\pm0.029$ $[0.05]$&$13.777\pm0.028$ $[-0.20]$\\
\morehorsp
$H_0$ [km/s/Mpc] &$70.3\pm1.6$&$67.03 \pm 0.61$ $[1.91]$&$67.84\pm0.72$ $[1.4]$&$68.25\pm0.69$ $[1.18]$\\
\morehorsp
$A_{lens}$&$1$&$1$ &$1.154\pm0.076$&$1.194\pm0.076$\\
\bottomrule
\end{tabular}
\end{center}
\caption{Constraints at $68 \%$ c.l. on cosmological parameters from pre-Planck datasets (second column, see \cite{calabresenew}), Planck TTTEEE in case of $\Lambda$CDM (third column), and Planck TTTEEE and Planck TTTEEE+WL varying $A_{lens}$ (fourth and fifth column, repectively). In the square brackets we report the shift S, defined via Eq.\eqref{Eq.shift}, that quantifies the discrepancy in the constraint on the parameter $\Pi$ between pre-Planck and Planck measurements. As we can see, when $A_{lens}$ is included, the tensions on the value of the Hubble constant, the matter and cosmological constants densities and the value of $\sigma_8$ are significantly reduced, especially when including cosmic shear data (WL).}
\label{tab:alens}
\end{table*}

In this section we discuss the current status of the $A_{lens}$ anomaly and its impact on current cosmological parameter estimation. In Table~\ref{tab:alens} we compare the constraints presented in \cite{calabresenew} with those derived from Planck 2015 temperature and polarization data assuming $\Lambda$CDM (third column) and a variation in $A_{lens}$ (see fourth column of the table). We aso show the effects of including cosmic shear data from CFHTLenS (named WL) as in \cite{planck2015} (fifth column). In the square brackets, on the right side of the constraint, we also report the shift $S$ between the cosmological constraints from Planck and pre-Planck measurements defined as:
\begin{equation}\label{Eq.shift}
  	S={{|\Pi_{pre-Planck}-\Pi_{Planck}|}\over{\sqrt{\sigma^2_{pre-Planck}+\sigma^2_{Planck}}}}
\end{equation}

where $\Pi$ and $\sigma$ are the parameter mean value and uncertainty reported for the pre-Planck and Planck datasets. As we can see, the most relevant (at about $\sim 2 \sigma$) shifts on the values of $\Omega_{m}$, $\sigma_8$ and $H_0$ are relieved when a variation in $A_{lens}$ is considered, especially when also the WL dataset is included. As we can see, we obtain a value for $A_{lens}>1$ at about $2$ sigma level from Planck TTTEEE and at about $2.6$ sigma from Planck TTTEEE+WL. The inclusion of cosmic shear data therefore does not only improve the agreement with the WMAP constraints but also the statistical significance for $A_{lens}$. 

\section{Method}

\begin{table}
\begin{center}
\begin{tabular}{lccccc}
\toprule
\horsp
Experiment \vertsp Beam \vertsp Power noise $w^{-1}$\vertsp $\ell_{max}$& $\ell_{min}$& $f_{sky}$\\
&  &[\footnotesize$\mu$K-arcmin]& & &\\
\hline
\morehorsp
Pixie      & $96$' & $4.2$ & $500$&$2$&$0.7$\\
\morehorsp
LiteBIRD      & $30$' & $4.5$& $3000$&$2$&$0.7$\\
\morehorsp
CORE      & $6$' & $2.5$& $3000$&$2$&$0.7$\\
\morehorsp
CORE-ext      & $4$' & $1.5$& $3000$&$2$&$0.7$\\
\morehorsp
Stage-III (Deep)      & $1$' & $4$ & $3000$&$50$&$0.06$\\
\morehorsp
Stage-III (Wide)      & $1.4$' & $8$ & $3000$&$50$&$0.4$\\
\morehorsp
Stage-IV      & $3$' & $1$& $3000$&$5,50$&$0.4$\\
\bottomrule
\end{tabular}
\end{center}
\caption{Experimental specifications for the several configurations considered in the forecasts.}
\label{tab:spec}
\end{table}

The goal of this paper is to investigate to what extent future CMB experiments will be able to constrain the value of $A_{lens}$ and falsify/confirm the current anomaly.
We have therefore simulated CMB anisotropy and polarization angular spectra data with a noise given by:

\begin{equation}
N_\ell = w^{-1}\exp(\ell(\ell+1)\theta^2/8\ln2),
\end{equation}

\noindent where $w^{-1}$ is the experimental power noise expressed in $\mu K$-arcmin and $\theta$ is the experimental FWHM angular resolution. We have considered several future experiments with technical specifications listed in Table~\ref{tab:spec}.
In particular, we have considered three possible CMB satellite experiments as CORE \cite{core1,core2}, LiteBIRD \cite{litebird} and PIXIE \cite{pixie}. A Stage-III experiment in two possible configurations as in \cite{erminia}, i.e. a 'wide' experiment similar to AdvACT  and a 'deep' experiment similar to SPT-3G. Finally we consider the possibility of a 'Stage-IV' experiment  as in \cite{erminia} (but see also \cite{stage4,Capparelli:2017tyx}).

\begin{table}
\begin{center}
\begin{tabular}{lc}
\toprule
\horsp
Parameter \vertsp Value \\
\hline
\morehorsp
$\Omega_{b}h^2$&$0.02225$\\
\morehorsp
$\Omega_{c}h^2$&$0.1198$\\
\morehorsp
$\tau$&$0.055$\\
\morehorsp
$n_s$&$0.9645$\\
\morehorsp
$100\theta_{MC}$&$1.04077$\\
\morehorsp
$ln(10^{10}A_{s})$&$3.094$\\
\morehorsp
$A_{lens}$&$1.00$\\
\bottomrule
\end{tabular}
\end{center}
\caption{Cosmological Parameters assumed for the fiducial model.}
\label{tab:par}
\end{table}

We have computed the theoretical CMB angular power spectra $C_{\ell}^{TT}$, $C_{\ell}^{TE}$, $C_{\ell}^{EE}$, $C_{\ell}^{BB}$ for temperature, cross temperature-polarization and $E$ and $B$ modes polarization using the CAMB Boltzmann code ~\cite{camb}. The angular spectra are generated assuming a fiducial flat $\Lambda$CDM  model with parameters compatible with the recent Planck 2015 constraints \cite{plancknewtau}. 

The theoretical $C_{\ell}$'s are then compared with the simulations using the Monte Carlo Markow Chain code {\sc CosmoMC}\footnote{\tt http://cosmologist.info}~\cite{Lewis:2002ah} based on the Metropolis-Hastings algorithm. The convergence of the chains is verified by the Gelman and Rubin method. Given a simulated dataset, for each theoretical model we evaluate a likelihood ${\cal L}$ given by

\begin{equation}
 - 2 \ln {\cal L} = \sum_{l} (2l+1) f_{\rm sky} \left(
\frac{D}{|\bar{C}|} + \ln{\frac{|\bar{C}|}{|\hat{C}|}} - 3 \right),
\label{chieff}
	\end{equation}

where $\bar{C}_l$ are the fiducial spectra plus noise (i.e. our simulated dataset) while $\hat{C}_l$ are the theory spectra plus noise.
$|\bar{C}|$, $|\hat{C}|$ are given by:

\begin{eqnarray}
|\bar{C}| &=& \bar{C}_\ell^{TT}\bar{C}_\ell^{EE}\bar{C}_\ell^{BB} -
\left(\bar{C}_\ell^{TE}\right)^2\bar{C}_\ell^{BB} ~, \\
|\hat{C}| &=& \hat{C}_\ell^{TT}\hat{C}_\ell^{EE}\hat{C}_\ell^{BB} -
\left(\hat{C}_\ell^{TE}\right)^2\hat{C}_\ell^{BB}~,
\end{eqnarray}

with $D$ defined as
\begin{eqnarray}
D  &=&
\hat{C}_\ell^{TT}\bar{C}_\ell^{EE}\bar{C}_\ell^{BB} +
\bar{C}_\ell^{TT}\hat{C}_\ell^{EE}\bar{C}_\ell^{BB} +
\bar{C}_\ell^{TT}\bar{C}_\ell^{EE}\hat{C}_\ell^{BB} \nonumber\\
&&- \bar{C}_\ell^{TE}\left(\bar{C}_\ell^{TE}\hat{C}_\ell^{BB} +
2\hat{C}_\ell^{TE}\bar{C}_\ell^{BB} \right). \nonumber\\
\end{eqnarray}

In what follows we also test the possibility of a angular dependence for $A_{lens}$. Such scale dependence could arise from beyond standard model physics such as modified gravity, cold dark energy, or massive neutrinos. We therefore consider the following parametrization (see \cite{DiValentino:2017rcr}):

\begin{equation}
A_{lens}(\ell) = A_{lens,0}( 1 + B_{lens}*\log(\ell/\ell_{*}) )
\label{alensdep}
\end{equation}

\noindent considering also the parameters $A_{lens,0}$ and $B_{lens}$ as free parameters and different values of the pivot scale $\ell_{*}$.

\section{Results}

\subsection{Future constraints on $A_{lens}$}

\begin{table}
\begin{center}
\begin{tabular}{lcc}
\toprule
\horsp
Experiment                    &$A_{lens}$ \\
\hline
\morehorsp
Pixie      & $1.016^{+0.09}_{-0.11}$\\
\morehorsp
LiteBIRD      & $1.001\pm0.025$\\
\morehorsp
CORE      & $1.001\pm0.013$\\
\morehorsp
CORE-ext      & $1.002\pm0.011$\\
\morehorsp
Stage-III (deep) & $0.92^{+0.13}_{-0.11}$\\
\morehorsp
Stage-III (wide)  &$0.97^{+0.11}_{-0.07}$\\    
\morehorsp
Stage-III (deep)+$\tau$-prior& $1.004^{+0.044}_{-0.048}$\\
\morehorsp
Stage-III (wide)+$\tau$-prior& $1.001^{+0.026}_{-0.028}$\\
\morehorsp
Stage-IV ($l_{min}=50$)    & $0.998\pm0.025$\\
\morehorsp
Stage-IV   ($l_{min}=5$)  & $0.999\pm0.015$\\
\bottomrule
\end{tabular}
\end{center}
\caption{Expected constraints on $A_{lens}$. The fiducial model assumes $A_{lens}=1.000$. For Stage-III wide, deep and Stage-IV with $l_{min}=50$ we have further imposed a Gaussian prior on the reionization optical depth corresponding to Planck 2015 results : $\tau = 0.055 \pm 0.010$.}
\label{tab:results}
\end{table}

The expected constraints on $A_{lens}$ for several future CMB experiments are reported in Table~\ref{tab:results}. As we can see a satellite experiment as PIXIE, devoted mainly to the measurement of CMB spectral distortions, will not have enough angular resolution to constrain $A_{lens}$, conversely a satellite as LiteBIRD, despite the poorer angular resolution with respect to Planck, thanks to the precise measurement of CMB polarization, could reach an accuracy of $\Delta A_{lens} \sim 0.026$, enough to falsify the current value of $A_{lens}\sim 0.15$ at more than five standard deviations.
A more ambitious space-based experiment as CORE, on the other hand, could test the $A_{lens}$ anomaly at more than $10$ standard deviations.
Near future ground-based as Stage-III will not have enough sensitivity on $A_{lens}$ unless the optical depth can be complementary measured by a different experiment. As we can see, considering an external prior on the optical depth as $\tau=0.055\pm0.010$ (in agreement with the recent Planck constraint \cite{plancknewtau}) can improve the Stage-III (Deep) constraint to a level comparable with LiteBIRD, while Stage-III (Wide) can also improve but with an accuracy smaller by about a factor two. A Stage-IV experiment can measure $A_{lens}$ with an accuracy about a factor $\sim 4.5$ better than the current Planck constraint, providing a large angular scale sensitivity from $l_{min}=5$. In this case, the current indication for $A_{lens}\sim1.15$ can be tested by a Stage=IV experiment at the level of $\sim 10$ standard deviations.
In the less optimistic case of a smaller sensitivity from $l_{min}=50$, the Stage-IV experiment is expected to constrain the $A_{lens}$ parameter with a precision comparable with the one achievable by LiteBIRD.

\subsection{Testing $A_{lens}$ in different spectra and frequency channels}

There are two, straightforward, ways for testing if the $A_{lens}$ anomaly is due to a systematic in the data: checking for its presence in the temperature and polarization spectra separately and considering also the frequency dependence. Of course, if the $A_{lens}$ anomaly is not simultaneously present in all the spectra and at all the frequencies this could better support the hypothesis of a systematic or a unresolved foreground. However when analyzing just one $C_{\ell}$ spectrum or just one frequency at time, the experimental noise is clearly larger and it is therefore interesting to investigate what kind of accuracy could be reached in this case.

As an example, we have considered the optimistic CMB-S4 configuration and considered the constraints on $A_{lens}$ achievable when using just the $TT$ and $EE$ channels.
We have found the following constraints at $68 \%$ C.L.: $A_{lens} =1.000 \pm 0.044$ ($TT$) and $A_{lens}=1.000 \pm 0.024$ (from $EE$). So, in practice, E polarization data alone from CMB-S4 could test the current $A_{lens}\sim 1.15$ anomaly at the level of $5$ standard deviations.

A complete configuration for the CMB-S4 experiment is clearly not yet finalized. In order to study the frequency sensitivity to $A_{lens}$ we have however assumed three channels at $90$, $150$ and $220$ GHz with angular resolutions of $5$, $3$, and $2$ arcminutes and detector sensitivities of $2.2$, $1.3$ and $2.2$ $\mu K$arcmin respectively. We have found  from $TT$ data the constraints
$A_{lens}=1.003^{+0.044}_{-0.045}$, $A_{lens}=1.002^{+0.041}_{-0.045}$, and $A_{lens}=1.003^{+0.041}_{-0.046}$ for the $90$, $150$ and $220$ GHz channels respectively. Using the $EE$ data we have 
$A_{lens}=1.003^{+0.028}_{-0.028}$, $A_{lens}=1.002^{+0.023}_{-0.025}$, and $A_{lens}=1.003^{+0.023}_{-0.025}$ again for the $90$, $150$ and $220$ GHz channels respectively.

\begin{figure}[!hbt]
%\centering
\includegraphics[width=.48\textwidth]{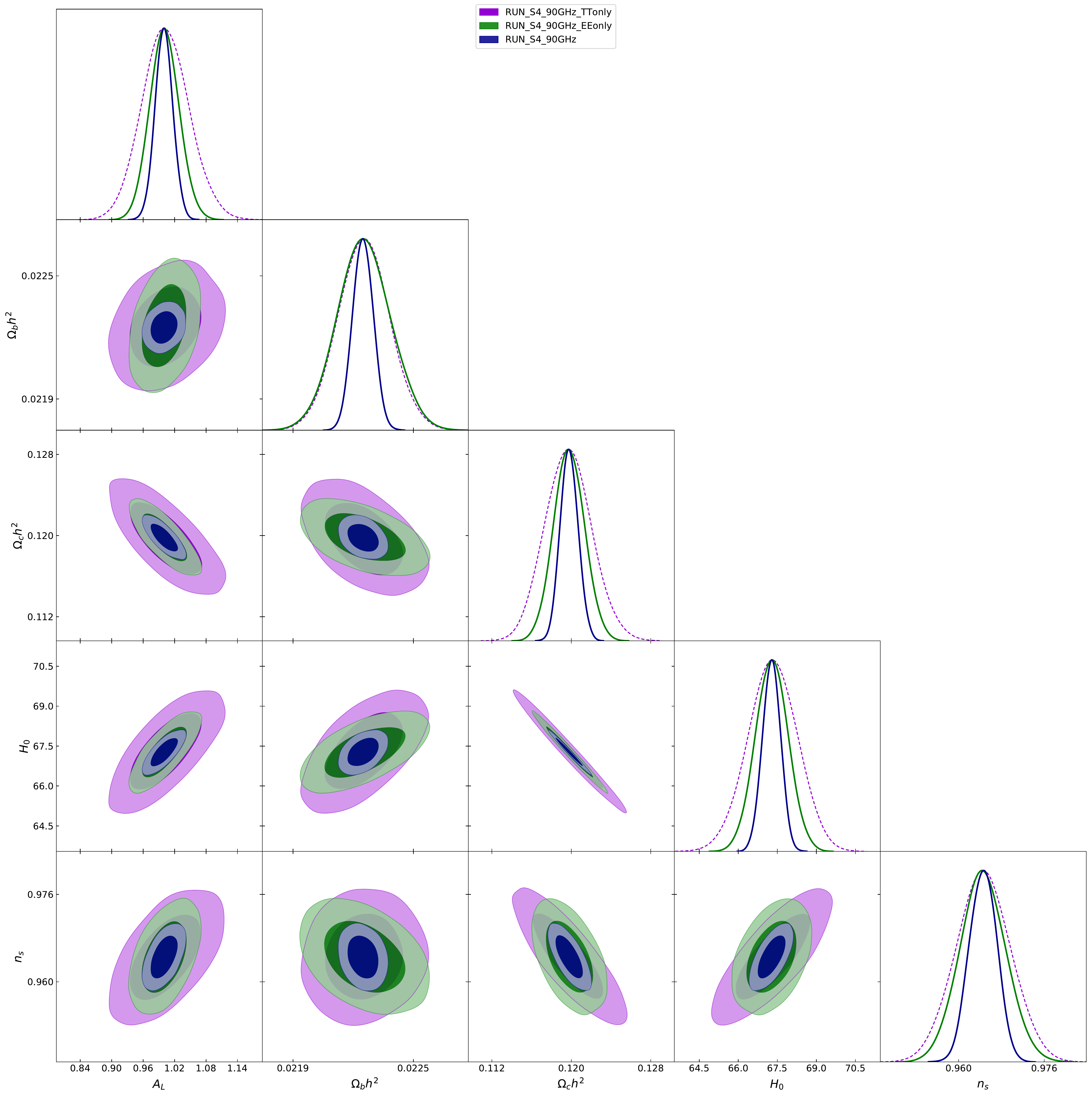}
\caption{Forecasted constraints at $68 \%$ and $95 \%$ C.L. for $A_{lens}$ and other cosmological parameters from a future CMB-S4 mission considering only the frequency channel at $90$ GHz.}
\label{figure90}
\end{figure}

\begin{figure}[!hbt]
%\centering
\includegraphics[width=.48\textwidth]{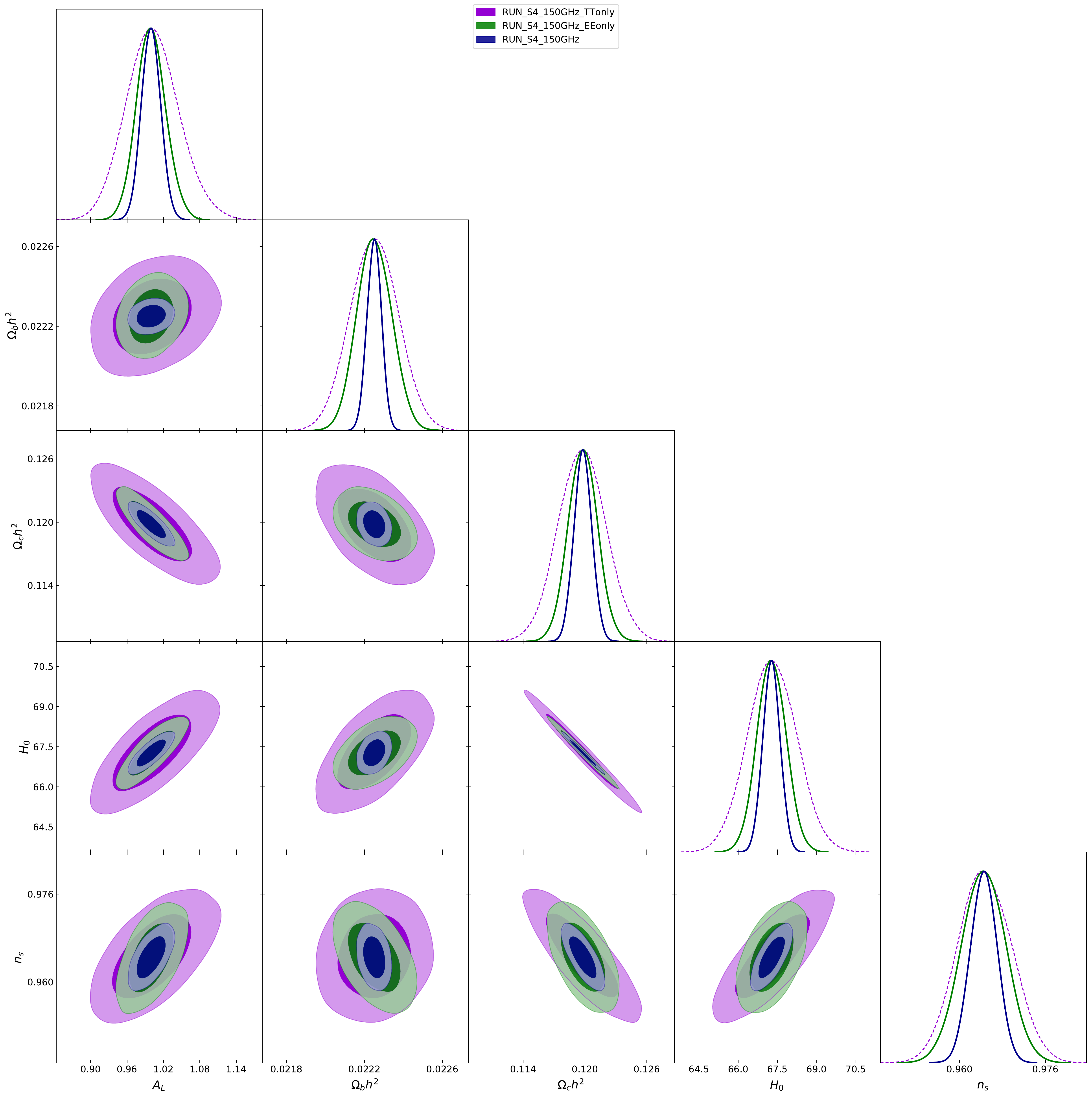}
\caption{Forecasted constraints at $68 \%$ and $95 \%$ C.L. for $A_{lens}$ and other cosmological parameters from a future CMB-S4 mission considering only the frequency channel at $150$ GHz.}
\label{figure150}
\end{figure}

\begin{figure}[!hbt]
%\centering
\includegraphics[width=.48\textwidth]{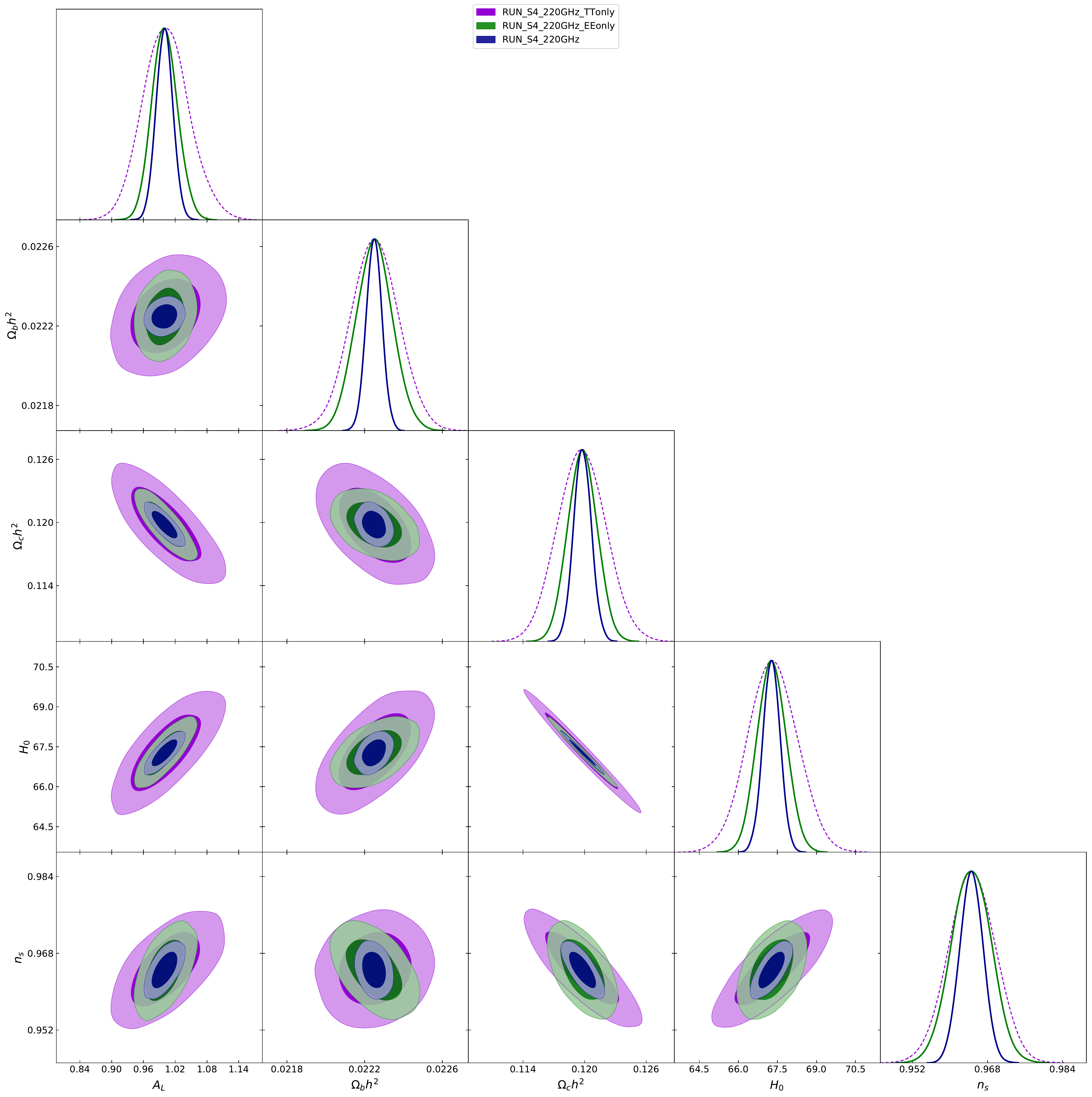}
\caption{Forecasted constraints at $68 \%$ and $95 \%$ C.L. for $A_{lens}$ and other cosmological parameters from a future CMB-S4 mission considering only the frequency channel at $220$ GHz.}
\label{figure220}
\end{figure}

In Figure ~\ref{figure90},\ref{figure150}, and \ref{figure220} we report the 2D forecasted constraints at $68 \%$ and $95 \%$ C.L. for $A_{lens}$ and other cosmological parameters from a future CMB-S4 mission considering the frequency channels at $90$, $150$, and $220$ GHz.

As we can see from the figures, polarization measurements will be crucial in improving the constraint on $A_{lens}$. In particular, polarization will somewhat reduce the degeneracy between $A_{lens}$ and the baryon density parameter present in TT data. However, $A_{lens}$ still strongly correlates with parameters as $n_S$, $\Omega_{cdm}h^2$, and $H_0$ even when the combined polarization+temperature measurements are considered.

As we can see, therefore, with the assumed experimental configuration, the sensitivity to $A_{lens}$ in each frequency channel will be essentially the same than the one achievable when all channels are combined. A frequency dependence of the $A_{lens}$ anomaly as a power law $\sim \nu^n$ could be tested with spectral indexes of $n\sim0.09$ at the level of three standard deviations.

\subsection{Using $B$ modes to test the $A_{lens}$ anomaly.}

\begin{figure}[!hbt]
%\centering
\includegraphics[width=.48\textwidth]{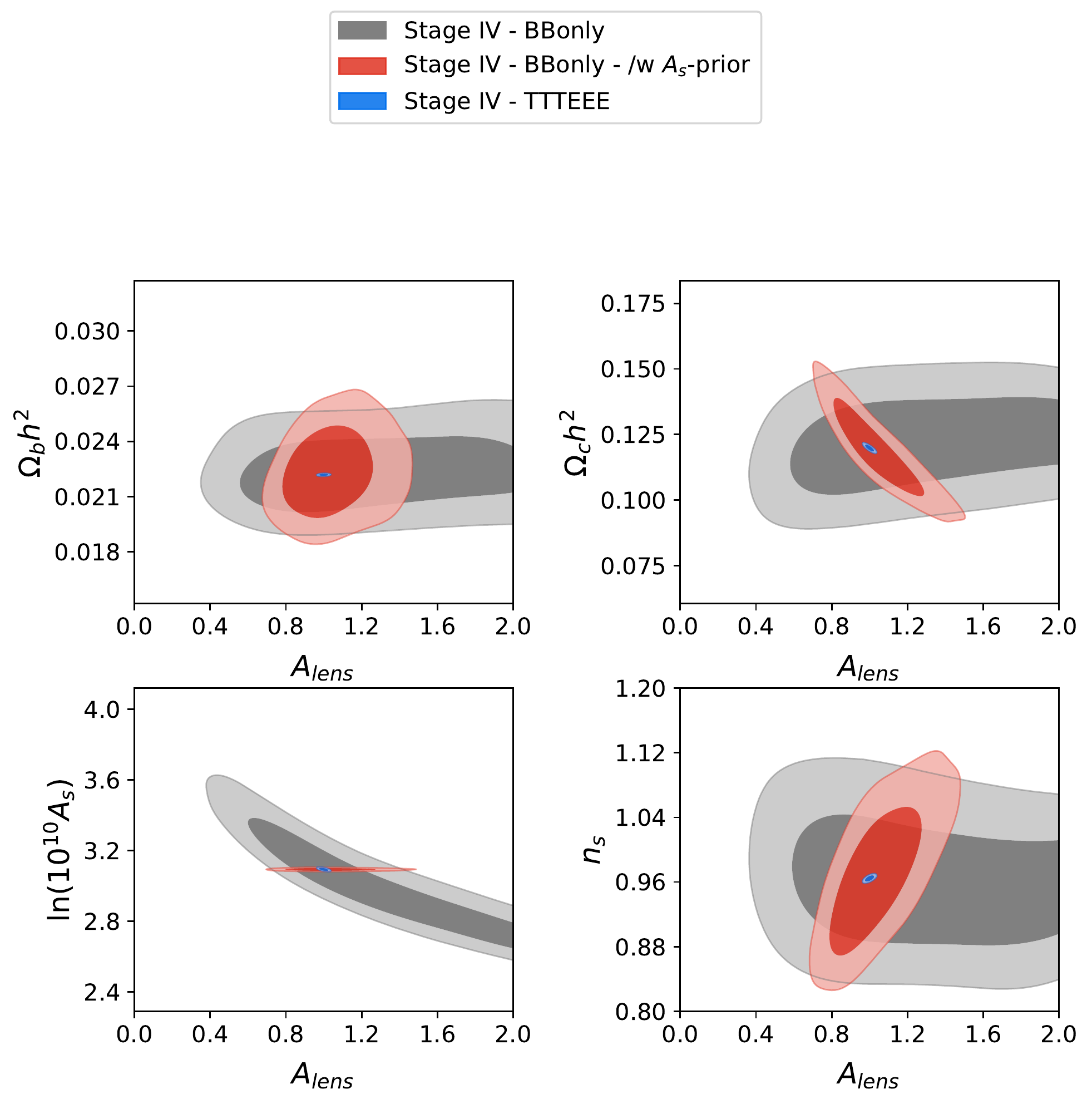}
	\caption{Future constraints at $68 \%$ and $95 \%$ C.L. from the Stage-IV experiment (with $l_{min}=5$) in the $A_{lens}$ vs $\Omega_bh^{2}$, $\Omega_ch^2$, $n_s$, and $ln[10^{10}A_s]$ planes (clockwise from Top Left panel). The constraints from BB modes only (Grey) leave $A_{lens}$ practically unbounded. Including a prior on the primordial amplitude improves the constraints on $A_{lens}$ from B modes only (Red) but they are still far weaker than the constraints from  TTTEEE (Blue).}
\label{figure1}
\end{figure}

Future experiments as Stage-IV will measure with great accuracy the CMB polarization B mode that arises from lensing. The $B$ mode spectra could therefore be in principle extremely useful for placing independent constraints on $A_{lens}$. In particular, an indication for an anomaly present in the TT, TE and EE angular spectra but not in the BB lensing spectrum would clearly confirm (once systematics or foregrounds are excluded) that the real physical nature of the anomaly is not connected to lensing but more to systematics or to new and unknowns processes possibly related to recombination or inflation that leave the small scale B mode signal as unaffected.
Unfortunately the polarization B mode signal does not only depends from $A_{lens}$. Degeneracies are indeed present between cosmological parameters and we have found that even with the Stage-IV experiment $A_{lens}$ will be practically unbounded from just the B mode spectra, with a major degeneracy with the amplitude of primordial perturbations $A_{s}$. 

Including an external Gaussian prior of $\log(10^{10}A_{s})=3.094\pm0.005$ for the primordial inflationary density perturbation amplitude and of $\tau = 0.055 \pm 0.010$ for the reionization optical depth, we found that Stage-IV could reach the constraint $A_{lens}=1.04^{+0.13}_{-0.19}$ at $68 \%$ c.l.. This would only marginally test the current anomaly and other complementary constraints will be needed to further test $A_{lens}$. In Figure I, we plot the future constraints at $68 \%$ and $95 \%$ C.L. from the Stage-IV experiment (with $l_{min}=5$) in the $A_{lens}$ vs $\Omega_bh^{2}$, $\Omega_ch^2$, $n_s$, and $ln[10^{10}A_s]$ planes. As we can see, the B modes are unable to bound $A_{lens}$ due mainly to a degeneracy with the primordial amplitude $A_s$. However, when a prior on $A_s$ is included, degeneracies are still present between $A_{lens}$ and $\Omega_bh^{2}$, $\Omega_ch^2$, and $n_s$ that prevent a precise determination of $A_{lens}$.

In conclusion, the measurement of primordial B modes from lensing will not let to significantly improve the constraints on $A_{lens}$ given the degeneracies between cosmological parameters.

\subsection{Future constraints on angular scale dependence of $A_{lens}$}

In Table~\ref{tab:results2} we report the constraints on the parameters of the angular scale dependency $A_{lens}$ in the form of Eq.\eqref{alensdep} for the Stage-IV configuration. For comparison, we also report the constraints using temperature and anisotropy spectra from the Planck 2015 release \cite{planck2015}.

As we can see, while the current bounds from Planck are rather weak and there is no indication for a scale dependency of the $A_{lens}$ anomaly (see also \cite{DiValentino:2017rcr}), the Stage-IV experiment can provide constraints at $\sim 1 \%$ level on $B_{lens}$, providing useful information on a possible scale dependence.
As discussed in the previous section, we have considered different pivot scales $\ell_*$. As we see from the results in Table~\ref{tab:results2}, while the choice of the pivot can change significantly current constraints, the effect on the accuracy  Stage-IV constraints is less significant.

\begin{table}[!hbtp]
\begin{center}
\begin{tabular}{ccc}
\toprule
\horsp
Parameter                   \vertsp Planck TTTEEE\vertsp Stage-IV \\
\hline
\hline
\morehorsp
$\ell_*$=50      & &\\
\morehorsp
$A_{lens,0}$      & $1.157^{+0.116}_{-0.144}$&$1.000\pm0.016$\\
\morehorsp
$B_{lens}$ & Unconstrained&$0.0002\pm0.0147$ \\
\hline
\hline
\morehorsp
$\ell_*$=300      & &\\
\morehorsp
$A_{lens,0}$      & $1.150^{+0.111}_{-0.139}$&$0.999\pm0.016$\\
\morehorsp
$B_{lens}$ & Unconstrained&$0.0002^{+0.0145}_{-0.0144}$ \\
\hline
\hline
\morehorsp
$\ell_*$=900      & &\\
\morehorsp
$A_{lens,0}$      & $1.220^{+0.181}_{-0.356}$&$0.999\pm0.019$\\
\morehorsp
$B_{lens}$ & Unconstrained&$-0.0004\pm0.0144$ \\
\hline
\hline
\morehorsp
$\ell_*$=1500      & &\\
\morehorsp
$A_{lens,0}$      & $1.269^{+0.209}_{-0.462}$&$0.999\pm0.021$\\
\morehorsp
$B_{lens}$ & Unconstrained&$-0.0005\pm0.0150$ \\
\hline
\hline
\morehorsp
$\ell_*$=2100      & &\\
\morehorsp
$A_{lens,0}$      & $1.313^{+0.223}_{-0.551}$&$0.999^{+0.022}_{-0.023}$\\
\morehorsp
$B_{lens}$ & Unconstrained&$-0.0004\pm0.0143$ \\
\bottomrule
\end{tabular}
\end{center}
\caption{Expected constraints on $A_{lens}$ and $B_{lens}$ from Planck real data and Stage-IV simulated data. The fiducial model for the simulated Stage-IV data has $A_{lens}=1.00$ and $B_{lens}=0.00$. We choose an hard flat prior $-0.4 < B_{lens} < 0.4$.}
\label{tab:results2}
\end{table}

\section{Conclusions}

While the agreement with the predictions of the $\Lambda$CDM model is impressive, the Planck data shows indications for a tension in the value of the  lensing amplitude $A_{lens}$ that clearly deserve further investigations. If future analyses of Planck data will confirm this tension then it will be the duty of new experiments to clarify the issue. In this brief paper we have shown that future proposed  satellite experiments as LiteBIRD can confirm/rule out the $A_{lens}$ tension at the level of $5$ standard deviation. The same accuracy can be reached by near future ground based experiments as Stage-III providing an accurate measurement of the reionization optical depth $\tau$ as already reported by Planck. Future, more optimistic, experiments as Stage-IV can falsify the $A_{lens}$ tension at the level of $10$ standard deviations. The Stage-IV experiment will also give significant information on the possible scale dependence of $A_{lens}$, clearly shedding more light on its physical nature.
A comparison between temperature and polarization measurements made at different frequencies could further identify possible systematics responsible for $A_{lens}>1$. We have shown that, in the case of the CMB-S4 experiment, polarization data alone will have the potential of falsifying the current $A_{lens}$ anomaly at more than five standard deviation and to strongly bound its frequency dependence.

\acknowledgments 
It is a pleasure to thank Eric Linder for useful comments and suggestions. EDV acknowledges support from the European Research Council in the form of a Consolidator Grant with number 681431. AM thanks the University of Manchester and the Jodrell Bank Center for Astrophysics for hospitality.

\end{document}